\documentclass[11pt,twoside]{article}
\usepackage{FUSE2004}
\usepackage{natbib}

\usepackage{epsf}
\usepackage{psfig}
\usepackage{lscape}

\begin{document}
\title{Phase-dependent changes in the wind lines in the HMXRB's SMC X-1 and 4U1700-37}
\author{R. C. Iping \& G. Sonneborn}
\affil{NASA's GSFC, code 681, Greenbelt MD 20771}

\author{L. Kaper \& G. Hammerschlag-Hensberge}
\affil{University of Amsterdam, Kruislaan 403, 1098 SJ Amsterdam}

\begin{abstract}
The High  Mass  X-ray Binaries (HMXRB's) SMC X-1 and  4U1700-37 have been observed with FUSE to study the effect of the X-ray source on the stellar wind of the primary.
In both systems phase dependent changes in the wind lines have been observed, indicating the creation of a X-ray ionization zone in the stellar wind. The high X-ray luminosity of SMC X-1 ionizes much of the wind and leaves a Stromgren zone. This disrupts the resonance-line acceleration of the wind in portions of the orbit, quencing the wind and disrupting the mass flow. A similar but less dramatic effect was found for the first time in 4U1700-37.  This so-called Hatchett-McCray (HM) effect had been predicted for 4U1700-37, but was not previously detected.  
\end{abstract}

\section{SMC X-1/Sk160}

 SMC X-1 is a luminous accretion-powered high-mass X-ray binary (HMXRB) with a B0 I primary and a rapid spinning pulsar (0.71 sec) in a 3.89-day period (Wojdowski et al. 1998). FUSE observed SMC X-1 for almost an entire 3.89 day orbital period in 2003 July. A total of 71 exposures (148 ksec) were obtained over 3.4 days, covering phase 0.20, through X-ray source conjunction ($\phi=0.5$), and primary eclipse ($\phi=0.93 - 0.07$). The exposures were divided into nine phase bins, each with $t_{expo}=14-20$ ksec and S/N$\sim15$ per 0.05 \AA.  The O\,VI 1032 stellar wind line profile was corrected for Milky Way and SMC interstellar H$_2$ and O\,VI 1032 absorption by simultaneously modelling the ISM lines in all nine spectra.  The resulting ISM model was subtracted from the observed O\,VI 1032 line at each phase (see Figure 1a for examples).  The equivalent width  of the O\,VI line was measured after the ISM correction.  Figure 1b shows the strong variation of the O\,VI equivalent width with orbital phase.
The orbital modulation of O\,VI in the stellar wind, the HM effect, is a result of ionization of the wind by the X-ray source.  The large X-ray luminosity of SMC X-1 ($L_x =2\times 10^{38}$ erg s$^{-1}$) ionizes much of the wind and leaves a large Stromgren zone.  This disrupts the resonance-line acceleration of the wind in portions of the orbit, quencing the wind and disrupting the mass flow. 
The O\,VI wind absorption is strongly asymmetric around the orbit. The line is at maximum strength during the eclipse of the pulsar $\phi=$ 0.0.  
The O\,VI line virtually disappears near $\phi \sim 0.4$. The O\,VI 1032 column density during eclipse is $\sim 7\times 10^{17}$ cm$^{-2}$. The O\,VI terminal velocity ($\sim -$700 km/s) drops to near zero at $\phi=$ 0.3-0.4. 

\begin{figure}[!ht]
\plotfiddle{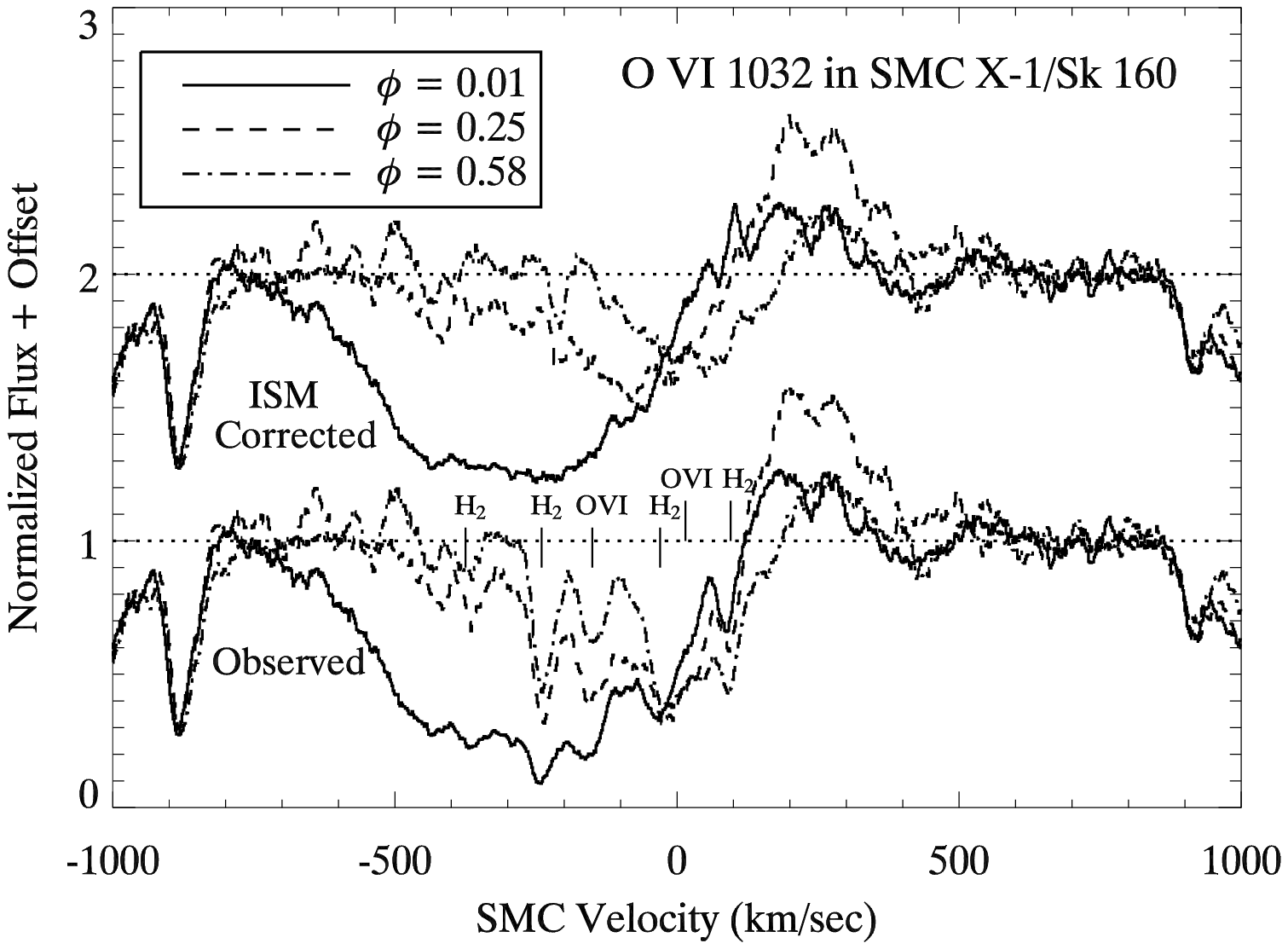}{1.0in}{0}{42}{42}{-240}{-240}
\plotfiddle{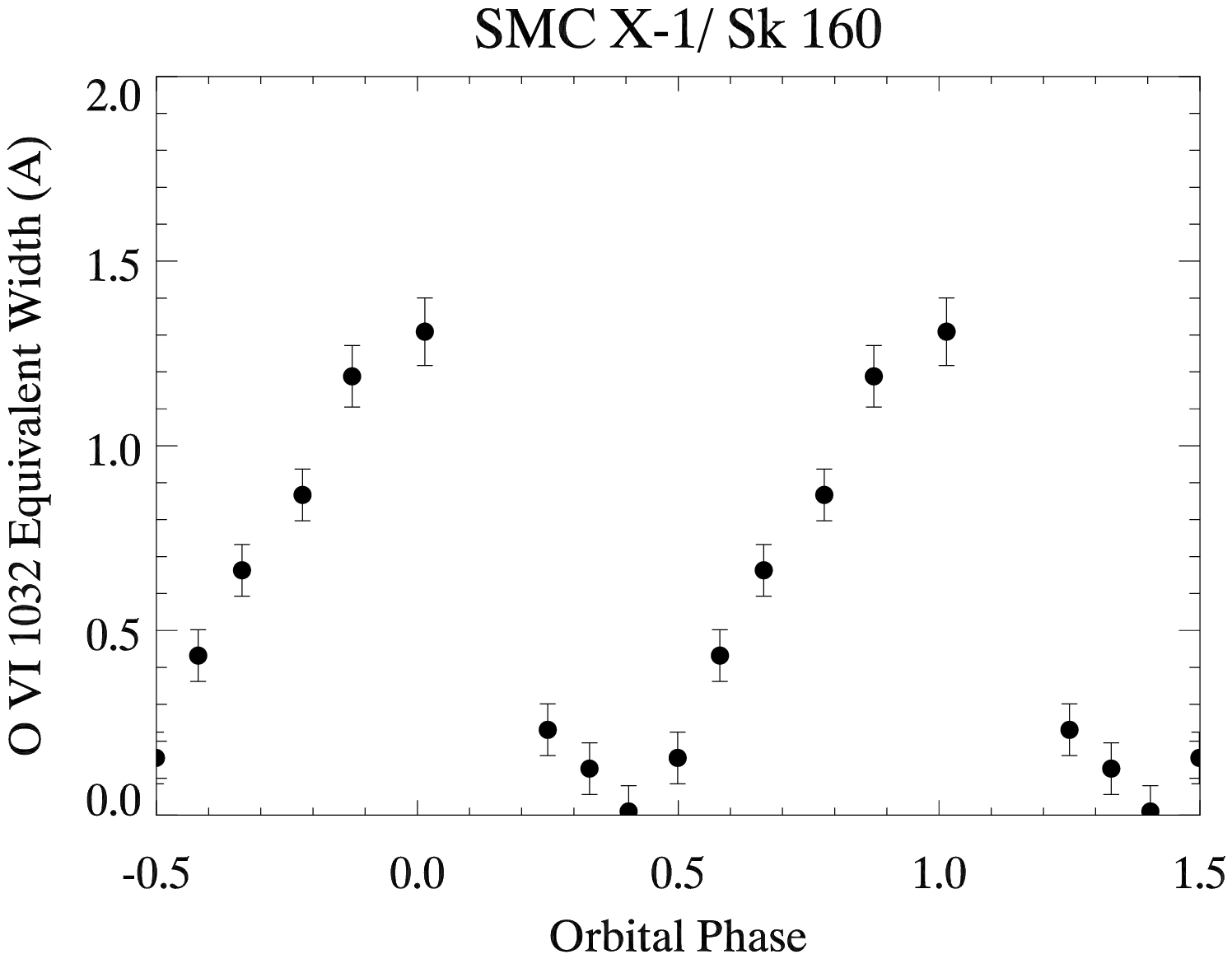}{1.0in}{0}{45}{45}{-50}{-170}
\caption{({\it left}) O\,VI 1032 stellar wind profiles are shown for three orbital phases (0.0, 0.25, and 0.58) of SMC X-1.  The observed profiles are in the lower panel, where the location of Milky Way and SMC interstellar H$_2$ and O\,VI features are identified.  The upper panel shows the same profiles after subtracting the ISM lines. ({\it right}) The equivalent width of the O\,VI 1032 stellar wind line is shown as a function of orbital phase.  The data points are repeated half a cycle before and after $0<\phi<1$ to illustrate the trend.}
\end{figure}

\section{4U1700-37/HD153919}
4U1700-37 has an O6.5 Iaf primary in a 3.451 day orbital period with a neutron star or black hole secondary. The system is believed to have escaped from Sco OB I about 2.5 million years ago. 
Most recent mass estimates by Clark et al. (2002), who found $M_1=58 M_{\odot}$ and $M_2=2.4 M_{\odot}$.
No X-ray /radio pulses or other periodicities are known, however RXTE ASM and CGRO BATSE data show evidence of a  13.8 day periodicity. (Hong \& Hailey, 2004). 4U1700-37 has been observed by FUSE at the four quadrature points of the binary orbit in 2003 April and August and the HM effect (Hatchett \& McCray 1977) has been observed for the first time in this system, 27 years after its prediction. The HM effect was predicted for 4U17000-37/HD153919, but was not detected in N\,V 1240, Si\,IV 1400, or C\,IV 1550 in IUE and HST data.
The P\,V 1118-1128 and S\,IV 1063-1073 P- Cygni lines are weakest at $\phi= 0.5$ (X-ray source conjuction) and strongest at $\phi= 0.75$ for the red wing and at $\phi= 0.0$ for the blue wing (see Figure 2).
The O\,VI and S\,VI wind lines show orbital modulation different from  P\,V and S\,IV and are strongest at $\phi= 0.5$ and weakest at $\phi= 0.0$ (X-ray source eclipse), implying that O\,VI and S\,VI are byproducts of the wind's ionization by the X-ray source. Such variations were not observed in N\,V, Si\,IV and C\,IV because of their high optical depth.  The P\,V and S\,IV transitions, on the other hand, are excellent tracers of the ionization conditions in the O star's wind.  P\,V is the dominant ionization stage in the wind and has lower cosmic abundance than C, N or Si.
There is very little change in the terminal velocity ($v \sim 2000 \pm 200$ km/sec) of the wind  with phase.

\begin{figure}[!ht]
\plotfiddle{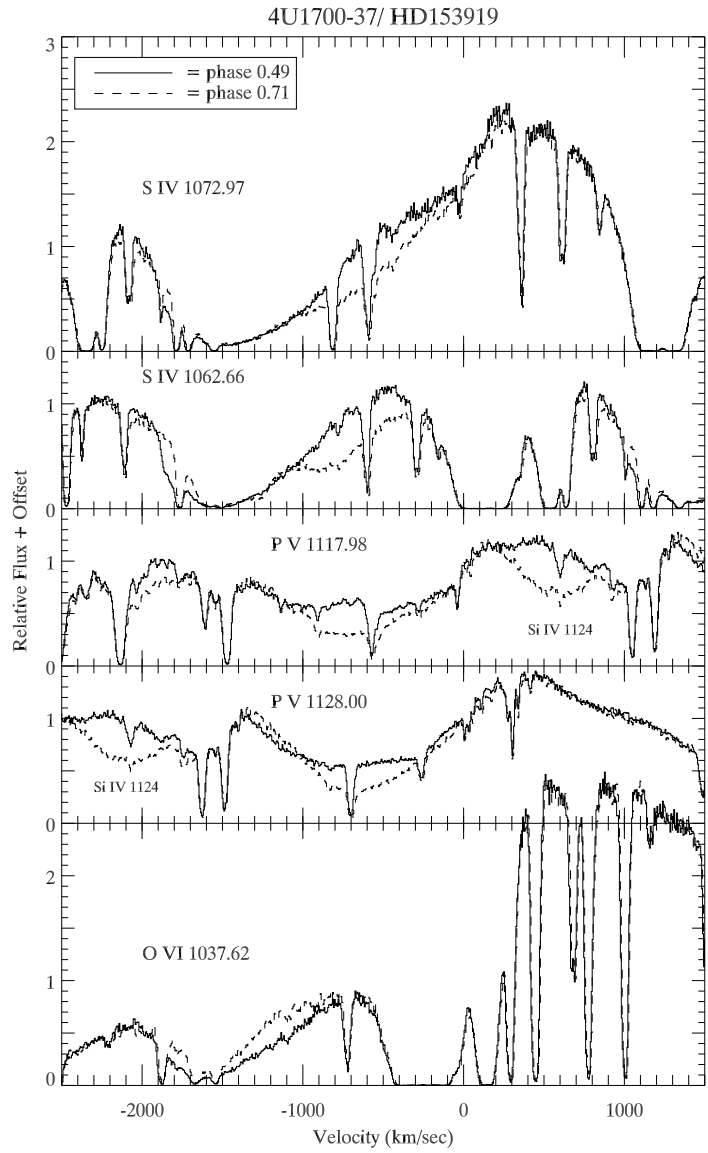}{5.9in}{0}{60}{60}{-200}{-20}
\caption{Stellar wind line profiles in 4U1700-37 at orbital phases 0.49 and 0.71.}
\end{figure}

\acknowledgements This work was supported in part by FUSE GI grant NNG04GK79G to Catholic University of America.



\begin{thebibliography}{}

\bibitem[]{}Clark, G., et al., 2002, A \& A 392, 909
\bibitem[]{}Hong, J., \& Hailey, C., 2004, ApJ, 600, 743
\bibitem[]{}Hatchett, S., \& McCray, R. 1977, ApJ, 211, 522
\bibitem[]{}Wojdowski, P., et al. 1998, ApJ, 502, 253
\end{thebibliography}


\end{document}